# Proposal of Automatic Offloading for Function Blocks of Applications

Yoji Yamato[a,*]

[a]Network Service Systems Laboratories, NTT Corporation, 3-9-11 Midori-cho, Musashino-shi, Tokyo 180-8585, Japan

*Corresponding Author: yoji.yamato.wa@hco.ntt.co.jp

## Abstract

When using heterogeneous hardware other than CPUs, barriers of technical skills such as OpenCL are high. Based on that, I have proposed environment adaptive software that enables automatic conversion, configuration, and high-performance operation of once written code, according to the hardware to be placed. Partly of the offloading to the GPU was automated previously. In this paper, I propose and evaluate an automatic extraction method of appropriate offload target loop statements of source code as the first step of offloading to FPGA. I evaluate the effectiveness of the proposed method in multiple applications.

## 1. Introduction

In recent years, it is said that Moore's Law will end shortly and CPU's density cannot be expected to double in 1.5 years. Based on this situation, systems with heterogeneous hardware such as GPU (Graphics Processing Unit) and FPGA (Field Programmable Gate Array) are increased. For example, Microsoft's search engine Bing tries to use FPGA [1]. AWS (Amazon Web Services) [2] provides GPU and FPGA using cloud technologies (e.g., [3]-[13]).

However, to achieve high performances using heterogeneous hardware for various applications, developers need to program and configure appropriately considering hardware and need to use expert technologies such as CUDA (Compute Unified Device Architecture) [14] and OpenCL (Open Computing Language) [15]. This is a high barrier to utilize GPU or FPGA.

Along with the progress of IoT (Internet of Things) technology (e.g., Industrie 4.0 and so on [16]-[21]), IoT devices are increasing rapidly, and many IoT applications are developed using service coordination technologies such as [22]-[30].

Expectation of applications utilizing heterogeneous hardware such as GPU and FPGA and many IoT devices is getting higher, however the hurdles are currently high for utilizing them. In order to break down such a situation, we think it is required in the future that application programmers only need to write logics to be processed, then software will adapt to the environments with heterogeneous hardware, to make it easy to utilize heterogeneous hardware and IoT devices.

Because Java [31] is insufficient for environment adaptation with performances, I have proposed environment adaptive software which run once written applications with high performance by automatically performing code conversion and configurations so that GPUs, FPGAs, IoT devices or so on can be used on deployment environments appropriately. As part of its technology, I have also realized automatic GPU or FPGA offloading of applications loop statements [32][33] partly. In this paper, I propose a method for offloading function blocks that are larger units rather than individual loop statement in applications to achieve higher performances by automatic offloading to GPU or FPGA. I implement the proposed method and evaluate the effectiveness of function block offloading using plural applications.

## 2. Existing Technologies

For GPGPU (General Purpose GPU) that uses GPU computational power not only for graphics processing (e.g., [34]), CUDA is a major development environment. To control heterogeneous hardware such as GPUs, FPGAs, and many core CPUs uniformly, OpenCL specification and its SDK (e.g., [35][36]) are widely used. CUDA and OpenCL need not only C language extension but also additional descriptions such as memory copy between GPU or FPGA devices and CPUs. Because of these programming difficulties, there are few CUDA and OpenCL programmers.

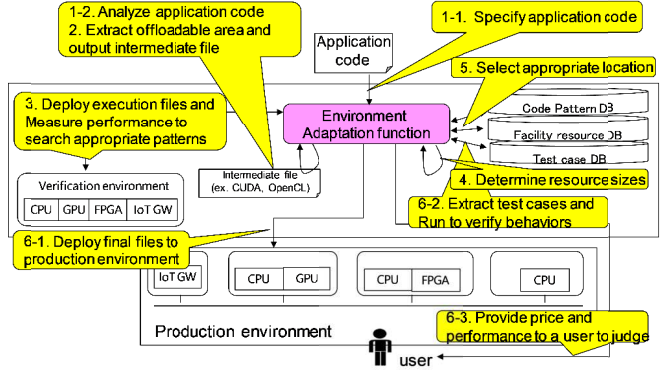

Fig. 1. Processing flow of environment adaptive software

For easy heterogeneous hardware programming, there are technologies that specify parallel processing areas by specified directives and compilers transform these directives into device oriented codes on the basis of specified directives. OpenACC [37] is one of the directive-based specifications, and the PGI compiler [38] is one of the directive-based compilers. For example, users specify OpenACC directives on C/C++ codes to process them in parallel, and the PGI compiler checks the possibility of parallel processing and outputs and deploys execution binary files to run on GPUs and CPUs. IBM JDK supports GPU offloading based on Java lambda expression [39].

In this way, CUDA, OpenCL, OpenACC, and others support GPU or FPGA offload processing. However, although processing on a GPU or FPGA itself can be performed, sufficient performance is hard to obtain. For example, when users use an automatic parallelization technology like the Intel compiler [40] for multicore CPU, possible areas of parallel processing such as "for" loop statements are extracted. However, naive parallel execution performances with GPUs or FPGAs are not high because of overheads of CPU and GPU/FPGA memory data transfer. To achieve high performances with GPU/FPGA, CUDA/OpenCL need to be tuned by highly skilled programmers or appropriate offloading area need to be searched for by the PGI compiler or others.

## 3. Proposal of Automatic Offloading for Function Blocks

### 3.1 Processing flow of environment adaptive software

In order to realize software adaptation to environment, I have proposed the following processing flow of environment adaptive software with reference to Figure 1. The environment adaptive software is realized in cooperation with functions including an environment adaptation function, a test case DB (using [41][42] and so on), a code pattern DB, a facility resource DB, a verification environment and a production environment.

### 3.2 Necessity of function blocks offloading

Firstly, I explain my previous automatic GPU offloading method for loop statements.

As a basic problem, it is possible for compilers to find the restriction that this loop statement cannot be processed in parallel with GPU, but it is difficult to find the suitability that this loop statement is suitable for GPU parallel processing. Therefore, an instruction to offload this loop to GPU is given manually, and performance measurements are repeated by trial and error.

Based on this situation, the paper of [32] proposes that GA [43] automatically finds an appropriate loop statement to be offloaded to GPU. First, a parallel loop statement is checked from a general purpose program that is not supposed to be parallelized, and loop statements offload patterns are mapped to genes with a value of 1 is set for GPU execution and 0 for CPU execution. Then, the performance verification trials are repeated in the verification environment to search for an appropriate offloading area.

Secondly, I explain my previous automatic FPGA offloading method for loop statements.

Even in FPGA, it is difficult to predict which loops will be faster when specific loop statements that take a long time to process are offloaded to FPGA. Therefore, we also propose to perform trial and error automatically in a verification environment like GPU cases. However, unlike GPU, since FPGA takes more than several hours to compile, we try actual FPGA measurements after narrowing down the offload candidate loop statements. For narrowed-down loop statements, our method generates OpenCL codes that offload each loop statement or combination of those loop statements, compiles them to FPGA, measures performances and selects the highest performance OpenCL code as the solution.

However, especially in the case of FPGA's acceleration, it is often the case that an algorithm for CPU is changed to an algorithm suitable for hardware processing. For this reason, simple offloading of loop statements were often

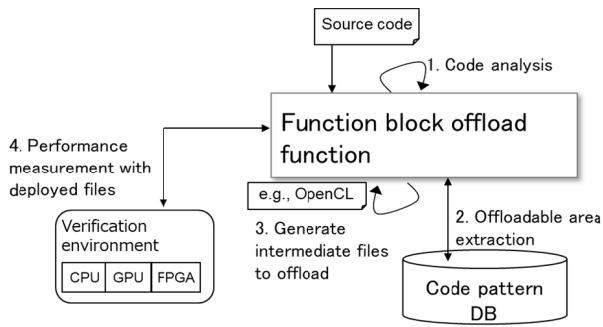

Fig. 2. Image of function blocks offloading.

insufficient in performance compared to improvements by manually changing algorithms.

It is currently very difficult for machines automatically extract hardware oriented algorithms for each application. Therefore, we aim to improve performance by replacing function blocks implemented with hardware oriented algorithms such as FPGA and GPU for large units such as matrix manipulation and Fourier transform in CPU codes. In other words, we use existing know-how of developers..

## 3.3 Outline of function blocks offloading and consideration points

Regarding to FPGA, because hardware circuit designs take a lot of time, it is often possible to use circuit design in the form of "IP core" for functions once designed.

As for GPU, FFT, linear algebra and image processing (e.g., [44]) are typical examples, and cuFFT, cuBLAS are implemented by CUDA and are provided free as GPU libraries. We consider using these libraries (not IP core) for GPU.

In this paper, if existing source code created for CPU includes function blocks that can be accelerated by offloading to GPU or FPGA such as FFT processing, GPU libraries or FPGA IP cores are replaced to the function blocks to speed up.

An overview of function blocks offloading is described in Figure 2. In Step 1, source codes are analyzed using a parse tool such as Clang, outer library calls and function processing are analyzed with loop statement structure. For library calls and function processing analyzed in Step 1, in Step 2, function blocks that can be offloaded to GPU or FPGA are found by checking with the code pattern DB. In Step 3, offloadable function blocks are replaced with libraries for GPU or IP cores for FPGA with creating interfaces with CPU programs. At this time, since it is not known whether function blocks offloading to GPU or FPGA will lead to immediate speedup, performance measurements are repeated in a verification environment to extract faster offloading patterns with or without certain function blocks offloading.

With regard to function block offloading, we need to consider following three points. Discovering function blocks in source codes, Checking whether the function blocks have offloadable GPU libraries or FPGA IP cores, Matching interfaces between replaced libraries or IP cores and host CPU program.

## 3.4 Function blocks offloading method

Based on three consideration points in the previous subsection, I study a function blocks offloading method in detail.

A. Discovering function blocks in source codes

A-1: In parsing, our method detects that external libraries are called from source codes. It is assumed that library calls such as FFT are detected. To detect them, the code pattern DB holds external libraries list and our method checks with the DB.

A-2 : In order to detect function processing other than registered library calls, classes, structures are detected from source code definition description by parse tools.

B. Checking whether the function blocks have offloadable GPU libraries or FPGA IP cores

B-1 : The code pattern DB holds GPU libraries, FPGA IP cores and related information which improve specific libraries or function block processing. For replacement source libraries and function blocks, codes and executable files with function names are registered. For library calls detected in A-1, our method searches for GPU libraries or FPGA IP cores that can be accelerated using library name as a key.

B-2 : The information registered in the code pattern DB in B-1 is used. The similarity detection tool detects whether there are libraries or IP cores that can be accelerated for the function processing of classes, structures detected in A-2. The similarity detection tool is a tool such as Deckard that detects a copy code or a changed code after copying. The similarity detection tool can detect some codes that are similar descriptions when calculated by CPU such as matrix manipulation, and changed descriptions after copying from other codes. The similarity detection tool cannot detect newly created classes, thus those are out of scope. For functions with libraries or IP cores registered in the code pattern DB that accelerate specific function blocks, the similarity detection tool judges the similarity is high or not based on the tool threshold.

C. Matching interfaces between replaced libraries or IP cores and host CPU program

C-1 : Since the corresponding library or IP core is searched in B-1 for the library call detected in A-1, the replacement library or IP core is installed in GPU or FPGA, and a host (CPU) program is connected. In the case of a library for GPU, a library such as CUDA is assumed. Since methods of using CUDA libraries from C language codes are open together with libraries, the code pattern DB holds library usage methods as well. When GPU libraries are used, GPU libraries and host program are connected referring to usage methods. In the case of an FPGA IP core, HDL is assumed. The code pattern DB also holds OpenCL code as IP core related information. From OpenCL code, the connection between CPU and FPGA using OpenCL interface and implementation of IP core on FPGA can be done via high-level synthesis tools of FPGA vendors such as Xilinx and Intel (Xilinx Vivado, Intel HLS Compiler, etc.).

C-2 : For classes and structures detected in A-2, we search for libraries and IP cores that can be accelerated in B-2, and we implement the corresponding libraries and IP cores on GPU and FPGA. Since B-2 is judged based on similarity, there is no guarantee that the basic parts such as the number and type of arguments and return match.

If they do not match, because libraries and IP cores are existing know-how and cannot be changed frequently, we will confirm a user whether to change according to the libraries or IP cores, and after receiving the confirmation, we will proceed performance tests.

## 4. Implementation

### 4.1 Tools to use

In this section, I explain the implementation of the proposed method. To confirm the method effectiveness of function blocks offloading, we use C/C++ language applications and NVIDIA Quadro P4000 (CUDA core: 1792, Memory: GDDR5 8GB) for GPU, Intel PAC with Intel Arria10 GX FPGA for FPGA.

GPU processing uses PGI compiler 19.4 in the market. PGI compiler is OpenACC compiler for C/C++/Fortran languages. PGI compiler also can use CUDA libraries such as cuFFT or cuBLAS.

To control FPGA, we use Intel Acceleration Stack Version 1.2 (Intel FPGA SDK for OpenCL 17.1.1, Quartus Prime Version 17.1.1). By including existing OpenCL codes of FPGA into kernel codes, those can be offloaded to FPGA during OpenCL program processing.

MySQL8.0 is used for the code pattern DB. It holds records for searching GPU libraries and FPGA IP cores that can be accelerated using the name of the library being called as a key. Libraries and IP cores have names, codes, and executable files associated with them. The usage method of the executable file is also registered. At the same time, code for comparison to detect the libraries and IP cores with the similarity detection tool is also held to associate with libraries and IP cores.

Deckard v2.0 [45] is used as the similarity detection tool. Deckard is used to expand function blocks for offloading. It judges the similarity between the partial code to be verified and the code for comparison registered in the code pattern DB to detect functions which are copied from outer files and changed. .

We implement the implementation by C language.

### 4.2 Implementation behavior

When a C/C++ application is specified, this implementation parses C/C++ code and detects loop statements and loop number for loop offloading of previous researches using gcov or gprof, called libraries (A-1) and defined classes and structures (A-2). For parsing, the Python program uses parsing libraries of LLVM/Clang [46] (libClang Python binding). When the implementation searches if there is an external library call, it checks the external library list in the code pattern DB.

Next, the implementation detects GPU libraries and FPGA IP cores that can speed up the library called (B-1). Using the called library name as a key, it obtains an executable file or OpenCL code that can be accelerated from the registered record in the code pattern DB. If a replacement function that can be accelerated is found, the implementation then generates an executable file. In the case of a GPU library, the implementation deletes the source part and replaces found GPU library call in the C/C++ code so that the replaced CUDA library is called. In the case of an IP core of FPGA, the implementation deletes the source part and replaces acquired OpenCL code to the kernel code. After completing the replacements, it compile with the PGI compiler for GPU and Intel Acceleration Stack for FPGA (C-1). For FPGA, based on OpenCL code, CPU and FPGA are connected via Intel's high-level synthesis tool.

Above description is the case of library call, detection processing is also performed in parallel when using similarity detection tool. The implementation uses Deckard to detect the similarity between the detected partial codes such as classes and the comparison code registered in the code pattern DB, and the comparison codes exceeding the threshold are detected (B-2). Detected codes are associated with corresponding GPU library or FPGA IP cores. Then, the implementation acquires executable files and OpenCL codes as same as B-1. Next, it generates executable files as same as C-1. However, if the interface of the source code and the replacement library or IP core arguments is different, the interface that matches the replacement library or IP core is notified to the user who requested the offload. The user can confirm whether it can be changed or not. And if the user accepts, the implementation generates executable files.

At this point, execution files are created that can be used to measure performances on GPU or FPGA in a verification environment. For function block offloading, if there is only one functional block to be replaced, we only consider whether the one is offloaded or not. However, if there are plural function blocks, the implementation generates a verification pattern that offloads a certain function block one by one to find a fast solution. This is because even if it is possible to increase the speed as existing know-how, it will not be clear whether the speed will be increased under the deployed environment condition until performance is actually measured. For example, if there are five function blocks that can be offloaded and the measurement results show that the performances of offloading of #2 and #4 can be improved, the implementation measures again with the pattern of offloading both #2 and #4. If it is faster than offloading with #2 and #4 alone, it selects both offloading as the solution.

## 5. Evaluation

### 5.1 Evaluation method

(a) Evaluated applications

I evaluate two applications, Fourier Transform and matrix calculation which are used many areas such as IoT.

The Fourier transform processing is used in various scenes of monitoring, such as vibration frequency analysis. When considering an IoT application that transfers data from a device to the network, it is assumed that the device side performs primary analysis such as FFT processing to reduce the network cost. In order to speed up FFT processing, the performance is improved by automatically replacing CUDA's existing library cuFFT [47].

Matrix calculation is used in many types of analysis such as machine learning analysis. Because matrix manipulation is used not only in cloud sides but also device sides along with spreading of IoT and AI, there are needs of automatic performance improvements for various applications including existing ones. In matrix calculation, LU decomposition processing of 2048 * 2048 orthogonal matrix data is performed. In order to speed up, the performance is improved by automatically replacing CUDA's existing library cuSOLVER [48].

For the Fourier transform and matrix calculation, the original codes are from Numerical Recipes in C [49].

(b) Experiment conditions

For function block offloading to GPU and FPGA, it is combined with loop statement offloading for actual use. However, since loop statement offloading has been evaluated previously in research [33] and so on, only GPU offload of function blocks is evaluated in this section. For the target applications, I prepare function blocks that can be offloaded in the code pattern DB beforehand and measure the performance when it is automatically replaced.

Conditions of experiments are as follows.

Offload source: Fourier transform application, Matrix calculation application

Offload target: cuFFT, cuSOLVER

Offload source discovery method: The code of the offload source application calls the external library on the code side and it is discovered by DB name matching. The application copies the library codes and puts comments and it is discovered by a similarity detection tool. I prepare both two patterns for verifications.

Methods to be compared: All CPU processing method, Proposed function block offloading method, Loop statement offloading method.

The loop statement offloading method is a work of [33] to search appropriate loop offloading patterns by Genetic Algorithm (GA) in a verification environment.

Performance measurement: In the Fourier transform, sample test processing is performed with the grid size is 2048 * 2048 and processing time is measured. In the matrix calculation, the processing time of LU decomposition for 2048 * 2048 orthogonal matrix is measured.

(c) Experiment environment

I use physical machines with NVIDIA Quadro P4000 for verifications. The CUDA core number of NVIDIA

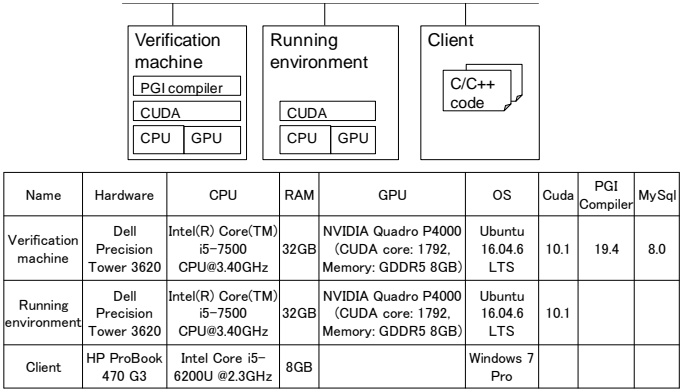

Fig. 3.  Experiment environment

Quadro P4000 is 1792. I use PGI compiler community edition v19.4 and CUDA toolkit v10.1. Figure 3 shows an experiment environment and environment specifications. Here, a client note PC specifies C/C++ application codes, codes are tuned with try and error on a verification machine, and final codes are deployed in a running environment for users after verifications.

**5.2  Performance results**

As applications that are expected to be used by many users in IoT and other areas, I confirmed performance improvements of Fourier transform and matrix calculation.

Figure 4 shows an example of Fourier transform performance improvement of previous research [33]. It shows maximum performance change of Fourier Transform in each generation with GA generation transitions (The vertical axis shows how many times faster GPU offloading was than using only CPU). Performances can be improved and GPU offloading is about 5.4 times faster.

Based on the previous results, I show the measurement results of how much performances have been improved by the proposed method implementation. First, the offload source discovery method can be replaced by the proposed DB name matching and similarity detection tool both whether the library is called or the code is copied. Figure 5 shows how many times the performances when function blocks offloading are performed compared to all CPU processing. 1 means the same performance of all CPU processing. The performance improvements of previous loop statement offloading [33] are also shown.

From Figure 5, it can be seen that the Fourier transform achieved 730 times performance by the proposed method, which was only 5.4 times in the previous loop statement offloading. As for matrix calculation, it was found that the proposed method has realized 130,000 times performance compared to 38 times in previous research. In previous research, it took more than a few hours to search for appropriate offloading loop statements using GA. However, the offloading of this function block has been completed in a few minutes.

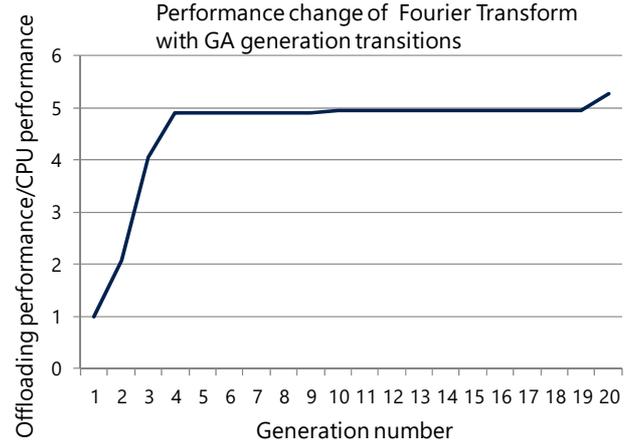

Fig. 4.  Reference graph: performance change of Fourier Transform with GA generation transitions [33]

|  | Performance improvement of loop offloading [33] | Performance improvement of function blocks offloading |
|---|---|---|
| Fourier transform | 5.4 | 730 |
| Matrix calculation | 38 | 130000 |

Fig. 5.  Comparison of performance improvement between loop offloading and proposed function block offloading

## 6.  Conclusions

In this paper, I proposed an automatic offloading method for function blocks of applications as a new elemental technology of environment adaptive software.

The proposed function block offloading method starts with source code analysis. It analyzes the source code, detects offloadable library calls with DB check, and replaces them with the use of replaceable GPU libraries or FPGA IP cores registered in the DB. The performance is measured in the verification environment, including the functions of the replaced GPU and FPGA, and the pattern with the highest performance is taken as the solution. In source code analysis, to search for more replaceable function blocks, offloadable function blocks are also searched using similarity detection technology. Replacement and performance measurement are performed as the same method. However, even if it is determined that the function block can be replaced, if the interface is different, the user is asked whether it can be changed with the interface of the replaceable function.